\documentclass[preprint,superscriptaddress,aps]{revtex4}
\usepackage{amsfonts,amsmath,amssymb}
\usepackage{amsfonts}
\usepackage{graphics}
\usepackage{color}
\usepackage{xcolor}\usepackage{color}
\usepackage{graphicx}
\usepackage{epsf}

\newcommand{\bea}{\begin{eqnarray}}
\newcommand{\eea}{\end{eqnarray}}

\begin{document}

\title{Compact-like vortices in isotropic curved spacetime}
	
\author{F. C. E. Lima}
\email[]{E-mail: cleiton.estevao@fisica.ufc.br}
\affiliation{Departamento de F\'{i}sica, Universidade Federal do Cear\'{a}, 
Campus do Pici, Fortaleza - CE, 60455-760, Brazil.}

\author{C. A. S. Almeida}
\email[]{E-mail: carlos@fisica.ufc.br}
\affiliation{Departamento de F\'{i}sica, Universidade Federal do Cear\'{a}, 
Campus do Pici, Fortaleza - CE, 60455-760, Brazil.}

\begin{abstract}
We introduced a generalized Maxwell-Higgs model in a $(3+1)$ isotropic spacetime, and we found their stationary solutions using the BPS approach in curved spacetime. In order to investigate the compact-like vortices, we assume a particular choice for the generalization term. The model is controlled by a potential driven by a single real parameter that can be used to change the vortex solutions profile as they approach their bound values. Resembling some flat spacetime vortex solutions, our model tends to compress the vortices when the parameter $l$ increases. Through numerical analysis, we also show the energy behavior and the magnetic field of the model.\\
\noindent Keywords: Gauge field theories, Generalized Maxwell model, Entended classical solutions.
\end{abstract}

\maketitle
\thispagestyle{empty}
\newpage

\section{Introduction}
The topological structure known as vortex appears when we study the scalar field coupled to the gauge field in a spacetime of $(2+1)D$.  Generally,  both fields evolve in space obeying the local $U(1)$ symmetry. The study of these structures in the relativistic context is introduced in ref. \cite{Nilsen,Vega} where the gauge field, i. e., the Maxwell field, is coupled to the Higgs field. It is important to mention that it is possible to study these structures in a context in which the dynamics of the gauge field is governed by the Chern-Simons field \cite{Jackiw}. An application not far this theory arises in the condensed matter physics when these vortex structures provide an explanation for the absence of free quarks in the QCD vacuum \cite{Singh,Giacomo}.

The detection of vortices in a lattice of linear rising potential \cite{Creutz} and the magnetic flux tube configurations \cite{Haymaker} support the study of the topological theory of these structures in QCD. Furthermore, in the literature, it is proposed that dual superconductivity is responsible for giving rise to vortices. Consequently, the dual superconductivity is related to $U(1)$ symmetry associated to topological charge conservation \cite{Giacomo}. After the discussion of the stability of solutions in the seminal paper of Bogomol'nyi \cite{BPS}, the study of vortex structures has become more attractive \cite{sch,kimm1,Ghosh,Sales,dionisio,sour,LA,LPA}. 

In recent years, the investigation of generalized models in flat spacetime has been considered as a new perspective, giving further motivations for studying topological structures \cite{bazeia,casana2,lima1,lima10}. The first work in this kind of generalized model was published by Lee and Nam \cite{nam}. Right after, Bazeia \cite{bazeia2} solves the first-order Bogomol'nyi-type equations of the model, using a supersymmetric approach. 

On the other hand, in the realm of cosmology, the first models with non-canonical kinetic term were discussed in the context of inflation \cite{Picon}. The generalized models were studied in an attempt to solve the problem of cosmic coincidence \cite{Picon1}. It is interesting to remark that these non-canonical kinetic models, in general, can present different characteristics from the standard case. As a matter of fact, we can note that the generalization of kinetic terms can change the field dynamics in a striking way.

Defined as finite wavelength solitons, compactons differ from the solitary waves not only by their compact support. Actually, after a collision of two compactons,  a low-amplitude compacton-anticompacton pairs are created \cite{rose,aro}. Other interesting feature of compactons is that they can only interact gravitationally \cite{santos}, which has been applied in brane cosmology \cite{adam2}. Here it is important to mention that this gravitational interaction is only possible if compactons are separated by a vacuum. In other words, there is no non-trivial overlap between these configurations. Investigations of the interaction of these configurations are suitable for understanding open problems in physics, such as dark matter \cite{adam3}. Recently, compactons were studied in the context of hybrid branes \cite{bazeia3,diego1,diego2}. Also a numerical approach to continuously transforming kinks into compactons in the O(3)-sigma model was introduced \cite{lima1}.

Currently, the literature does not present much work on compactons in curved spacetime. Indeed, there are only few works in compactons applied to brane models as we just mentioned. This is our fundamental motivation to propose and study a model where compactons are provided by a generalized Maxwell-Higgs theory in curved spacetime.

The main objective of this work is to study topological structures similar to the so-called vortices generated by the self-dual configurations obtained from a generalization of the Abelian Maxwell-Higgs model in curved spacetime. First, we performed a consistent implementation of the Bogomol'nyi method for the model and obtained the respective generalized self-dual equations in $(3+1)D$. The development of the BPS formalism allowed us to assume the form of the generalization function for the study of topological structures.

This work is organized as following: we started introducing the Maxwell-Higgs model in an isotropic spacetime and we found their stationary solutions using the BPS approach in curved spacetime. Then, we present a particular case, which solutions are solved by a numerical approach. Finally, we present and discuss the results obtained in our work.

\section{Generalized Maxwell Vortices}
Seeking to investigate vortices-like solutions that present compacton behavior in the $(3+1)D$ curved spacetime, we propose a generalized Abelian Higgs model such that the coupling of the Higgs scalar with the background metric is obtained through an $R\phi^{2}$ term, where $R$ is the Ricci scalar. Namely, 
\begin{align}\nonumber
    S=&\int\, d^{4}x\, \sqrt{-g}\bigg[-\frac{1}{4}F_{\mu\nu}F^{\mu\nu}+\frac{H(|\phi|)}{2}D_{\mu}\phi^{a}D^{\mu}\phi^{a}-\frac{e^{2}}{8}\beta\phi^{2l}(\phi^{2l}-F^{2})^{2}+\\
    -&\frac{1-\beta}{2}B(x)\phi^{2}\bigg] \label{us},
\end{align}
where we introduced a $\phi^{8}$-like spontaneous symmetry breaking potential. As we will shown, the parameter $l$ will play a crucial role in the compact features of vortices. We generalize the kinetic term for the introduction of a function ${H(|\phi|)}$. Also $g$ denotes det$g _{\mu\nu}$, whereas, $\phi^{a}$ $(a=1,2)$ is a charged scalar field. The covariant derivative as defined as
\begin{equation}
D_{\mu}\phi^{a}=\partial_{\mu}\phi^{a}+e\varepsilon^{ab}A_{\mu}\phi^{b}.
\end{equation}

\begin{figure}[ht!]
    \centering
    \includegraphics[scale=0.6]{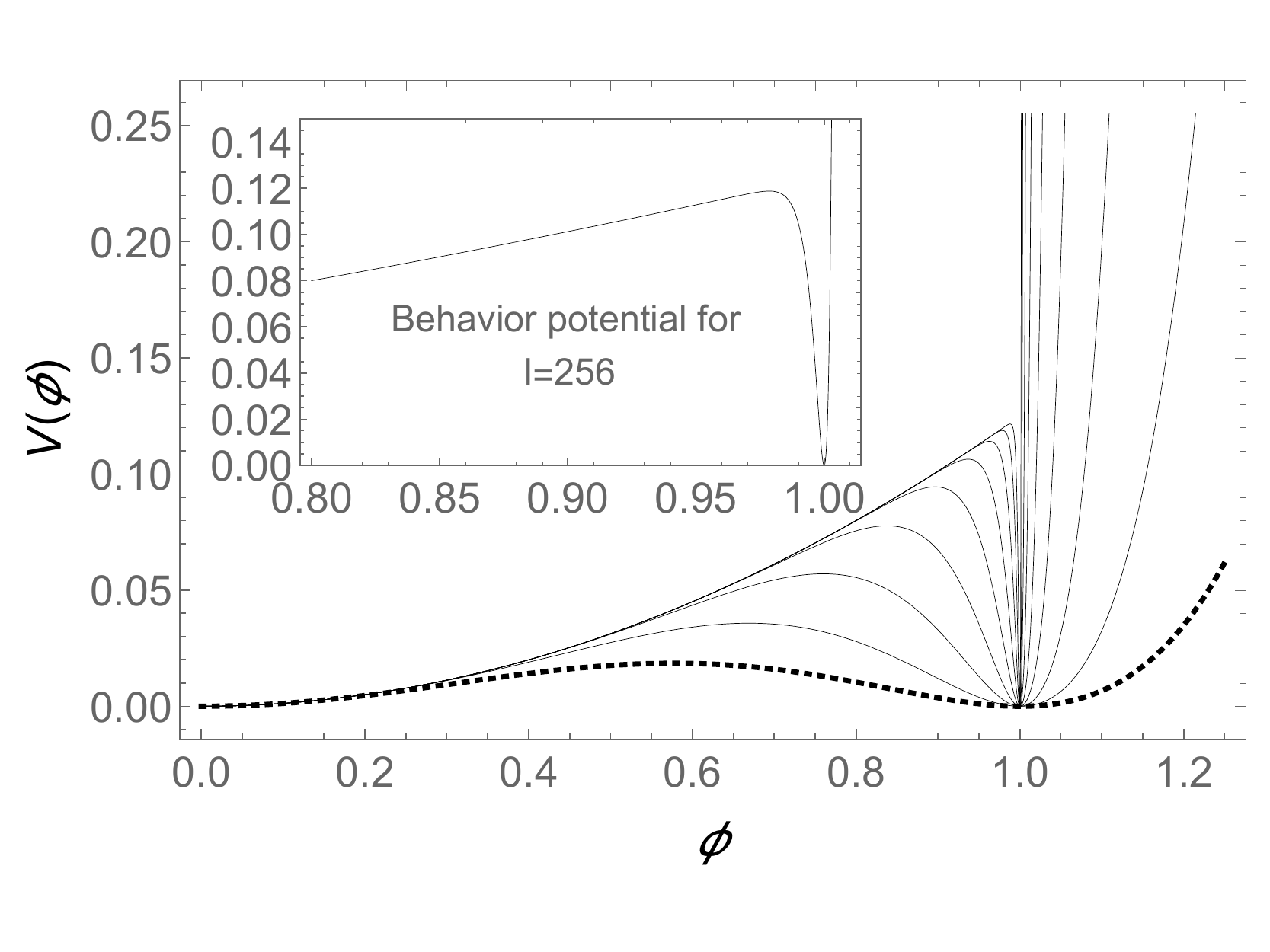}
    \vspace{-20pt}
    \caption{Potential associated with the action for several values of the parameter $l$. The values of $l$ are: 1, 2, 4, 8, 16, 32, 64, 128 and 256, with $l = 1$ the dotted line.}.
    \label{potential}
\end{figure}

The behavior of the potential for various values of $ l $ is shown in fig. \ref{potential}. This way, we note that as the parameter $l$ grows the physical potential becomes more localized around the vacuum state value of the model.

The scalar source $B(x)$ is introduced as a Lagrange multiplier in order to obtain later the coupling $R\phi^{2}$ and consistent BPS solutions.  The constant $\beta$ expresses a relationship between the usual symmetry breaking constant $\lambda$ and the gauge coupling constant $e$ in the form $\beta=\frac{8\lambda}{e^{2}}$. This model is a generalization of a model studied by Edelstein et. al. \cite{Edelstein}. Part of our calculations closely follow their developments.

As noted, the source term $ B (x) $ is related to a background metric, so that model admits the field settings with BPS properties. An interesting point of view is that the source term $B(x)$ can be interpreted as an external background field. Recently, there has been significant development in the topic of topological solitons with BPS properties in the presence of impurities. In this scenario, the background field $B(x)$ can also be understood as a frozen soliton kept in the background \cite{T,AQRW,S,NOB,ASHCRO}.


In this work our aim is to search for compact-like vortex solutions in a gauge model in the context of curved spacetime.
We will use the action (\ref{us}) as a starting point. 


The motion equation for the gauge field is 
\begin{equation}
    \frac{1}{\sqrt{-g}}\partial_{\mu}(\sqrt{-g}F^{\mu\nu})=-e\varepsilon^{ab}H\phi^{a}D^{\nu}\phi^{b},
\end{equation}
and for scalar fields is
\begin{align} \nonumber
    \frac{1}{\sqrt{-g}}D_{\mu}(\sqrt{-g}H D^{\mu}\phi^{a})=&2\lambda l\phi^{a}(\phi^{2l}-F^{2})^{2}+4\lambda l\phi^{a}\phi^{2l}(\phi^{2l}-F^{2})+(1-\beta)B\phi^{a} \\
    &+\frac{1}{2}H_{\phi^{a}}D_{\mu}\phi^{a}D^{\mu}\phi^{a},
\end{align}
where
\begin{equation}
    H_{\phi^{a}}=\frac{\partial H}{\partial\phi^{a}}.
\end{equation}

Now, we can describe the BPS equations to investigate the vortex structures, and we will consider an isotropic spacetime and symmetric. Therefore, we can use the following metric \cite{Edelstein,gibbons1,gibbons2}, 
\begin{equation}
    ds^{2}=V^{2}dt^{2}-Y^{2}dz^{2}-g_{ij}dx^{i}dx^{j},
\end{equation}
with $V$, $Y$ and $g_{ij}$ dependent only on the polar coordinate $r$. Assuming that
\begin{equation}
    g_{ij}=\frac{1}{V^{2}}\delta_{ij},
\end{equation}
the metric is reduced to
\begin{equation}
    ds^{2}=V^{2}dt^{2}-\frac{1}{V^{2}}(dx^{2}+dy^{2}+dz^{2}),
\end{equation}
where $Y^{2}=V^{-2}$. Here it is interesting to mention that we will consider the study of vortex configurations with axial symmetry, therefore the polar symmetry it is considered. In other words, we will use the notation: $(i,j)\equiv (r,\theta)$.

It is important to mention that the choice of the metric is related to the type of BPS equations that we want to obtain later in this work. We chose this type, to obtain vortex-like equations. Other choices of metrics may be not proper to be investigating vortices in this model. With the path chosen, it is possible to find new classes of topological structures.

Given the choice of the metric, we obtain that the scalar curvature is
\begin{equation}
    R=4(\partial V)^{2}-2V\partial^{2}V.
\end{equation}

Considering that the energy functional is obtained by integrating the component $T_{00}$ of the energy-momentum tensor and that this can be obtained by coupling the fields with gravity, then, by varying the action with respect to the metric, we obtain the expression
\begin{equation}
\label{TensorEM}
    T_{\mu\nu}=-\frac{1}{2}F_{\mu}^{\lambda}F_{\nu\lambda}+H D_{\mu}\phi^{a} D_{\nu}\phi^{a}-g_{\mu\nu}\mathcal{L},
\end{equation}
where as usual $A_{3}=0$.

The conserved current is 
\begin{equation}
    j^{\mu}=\sqrt{-g}T^{\mu}\,_{0},
\end{equation}
with an associated charge given by
\begin{equation}
    \mathcal{Q}=\int\, d^{3}x\, \sqrt{-g}j^{0}.
\end{equation}

Using the equation (\ref{TensorEM}), we have that the model energy is
\begin{equation}
\label{energy}
    E=\int\, d^{2}x \sqrt{-g}\bigg[\frac{1}{4}F_{ij}F_{ij}-\frac{1}{2}D^{i}\phi^{a}D_{i}\phi^{a}+\frac{e^{2}}{8}\beta\phi^{2l}(\phi^{2l}-F^{2})^{2}+\frac{1-\beta}{2}B(x)\phi^{2}\bigg],
\end{equation}
where the integrating of the above expression is the energy density.

Rearranging the expression of energy, we have that
\begin{align}\nonumber \label{E}
    E=&\int\, d^{2}x\bigg\{\frac{1}{4}\bigg[VF_{ij}\mp\frac{e\sqrt{\beta}}{2V}\varepsilon_{ij}\phi^{l}(\phi^{2l}-F^{2})\bigg]^{2}+ \frac{\beta}{4}V^{2(\beta-1)}\bigg[\varepsilon_{ij}V^{(1-\beta)}D_{j}\phi^{a}\\ \nonumber
    &\mp\sqrt{\frac{H}{\beta}}\varepsilon_{ab}D_{i}(V^{1-\beta}\phi)^{b}\bigg]^{2}+\frac{\beta(1-\beta)}{4}V^{-2}\bigg[\varepsilon_{ab}\sqrt{\frac{H}{\beta}}D_{i}(\phi V)^{b}\pm\varepsilon_{ij}\phi^{a}\partial_{j}V\bigg]^{2} \\ &+\frac{1-\beta}{4V^{2}}(2B-R)\phi^{2}+F^2\sqrt{\beta}\partial_i S_i\bigg\}.
\end{align}

The last term in eq. (\ref{E}) is the surface contribution of the topological current, namely,
\begin{align}
    S_{i}=\frac{1}{2}\varepsilon_{ij}(\varepsilon_{ab}\sqrt{H}\frac{\phi^a}{F}D_j\frac{\phi^b}{F}+eA_j),
\end{align}
in the case without generalization, i. e., for $H=1$ we get the result of Ref. \cite{Edelstein}, that is 
\begin{equation} 
    \oint S_i dx^i=\frac{2\pi}{e}n,
\end{equation}
with $n$ the topological charge of the vortex configuration.

In fact, it is important to mention that the last term of the expression (\ref{E}) is known as the BPS energy density of the model.

Note that the energy obeys the inequality, i. e., $E\geq E_{BPS}$. However, once the BPS bound is satisfied, we must have $E=E_{BPS}$. With this, we obtain the Bogomol'nyi equations for the model, namely:
\begin{equation}
\label{bps1}
    VF_{ij}=\pm\frac{e\sqrt{\beta}}{2V}\varepsilon_{ij}\phi^{l}(\phi^{2l}-F^{2}),
\end{equation}

\begin{equation}
\label{bps2}
    \varepsilon_{ij}V^{(1-\beta)}D_{j}\phi^{a}=\pm\sqrt{\frac{H}{\beta}}\varepsilon_{ab}D_{i}(V^{1-\beta}\phi)^{b},
\end{equation}

\begin{equation}
\label{bps3}
    \varepsilon_{ab}\sqrt{\frac{H}{\beta}}D_{i}(\phi V)^{b}=\mp\varepsilon_{ij}\phi^{a}\partial_{j}V.
\end{equation}

For the saturated energy of the model is saturated, or when we reach the BPS bound, the curvature tensor is intrinsically related to the source term. This coincides with the result in Ref. \cite{Edelstein}. Explicitly, we have that
\begin{equation}
    B=\frac{R}{2}.
\end{equation}

Note that the equation (\ref{bps1}) is the BPS equation that describes the dynamics of the gauge field in curved space. It is important to mention that the positive and negative signs have a correspondence with the topological charge of the vortex.

Analyzing the expressions (\ref{bps1}), (\ref{bps2}) and (\ref{bps3}), we realize that the equations (\ref{bps2}) and (\ref{bps3}) correspond to the same equation. In this way, we have only two independent differential equations.

Since we are studying a model with a generalization $H(|\phi|)$ coupled with the kinetic term, we obtain that the expression (\ref{bps3}) can be reduced to
\begin{equation}\label{lnV}
    \partial_{i}\text{ln}\phi^{2}=-2\partial_{i}\text{ln}V.
\end{equation}

Now, let us consider the asymptotic behaviors, namely
\begin{equation}
    \lim_{r\rightarrow\infty}V=1 \hspace{1cm} \text{and} \hspace{1cm} \lim_{r\rightarrow\infty}\vert\phi\vert=F^{1/l}.
\end{equation}

Then we can establish that 
\begin{equation}
\label{sol}
    \phi^{2}=\frac{F^{2/l}}{V^{2}}.
\end{equation}

For convenience, to study the field configurations, let us choose an ansatz for the scalar field of the shape
\begin{equation}
\label{ANSATZ}
    \phi=Fg(r)\text{e}^{i\alpha},
\end{equation}
with $g(r)$ and $\alpha$ two real functions. Thus, from the equation (\ref{bps2}) we have 
\begin{equation}
\label{sol1}
    \text{e}A_{j}=\partial_{j}\alpha\mp \frac{\sqrt{\beta H}}{2}\varepsilon_{ij}\partial_{j}\text{ln}g(r)^{2}.
\end{equation}

By replacing eq. (\ref{sol1}) in eq. (\ref{bps1}), we get that
\begin{equation}
\label{sol2}
    \partial^{2}\text{ln}g(r)^{2}+\frac{1}{2}\partial_{j}H\cdot\partial_{j}\text{ln}g(r)^{2}=\frac{\text{e}^{2}F^{2}}{V^{2}}g^{2l}(g^{2l}-1).
\end{equation}

The equation for the magnetic field can be written as
\begin{equation}
  \label{mag}
    \frac{\text{e}}{2}\varepsilon_{ij}F_{ij}=\pm\frac{m_{V}^{2}\sqrt{\beta}}{2}g^{2l}(g^{2l}-1)g^{2},
\end{equation}
with $m_{V}=\text{e}F$.

Using the ansatz of eq. (\ref{ANSATZ}), we have that the components of the scalar field for a configuration of $n$-vortices can be given by 
\begin{equation}
\label{ansatz1}
    \phi^{1}=Fg(r)\cos{n\theta},
\end{equation}
and
\begin{equation}
\label{ansatz2}
    \phi^{2}=Fg(r)\sin{n\theta}.
\end{equation}

We also consider that
\begin{equation}
    \label{ansatz3}
\text{e}A_{\phi}=-\frac{A(r)-n}{r}.
\end{equation}
Therefore, we should note that
\begin{equation}
    V^{2}=g(r)^{-2}.
\end{equation}

Finally, considering the ansatz for the components of the scalar field and for the gauge field and assuming that $A(r)/\sqrt{\beta}\equiv a(r)$, the expressions that describe the dynamics of the fields can be written as
\begin{equation}
    \label{eq1}
    rg'=\pm\sqrt{H(g)}\frac{a(r)g(r)}{r},
\end{equation}
and
\begin{equation}
    a'(r)=\mp\frac{m_{V}^{2}}{2}rg^{2(l+1)}(g^{2l}-1).
\end{equation}

Decoupling the Bogomol'nyi equations, we obtain that
\begin{equation}
    rg''(r)+2g'(r)-\frac{1}{\sqrt{H}}\frac{d\sqrt{H}}{dr}\frac{rg'(r)}{g(r)}+\frac{1}{2}m_{V}^{2}H^{1/2}g(r)^{2l}(g(r)^{2l}-1)g(r)^{3}-rg(r)=0.
\end{equation}

It is worthy to recall here the seminal paper of Jackiw and Weinberg \cite{Jackiw} about vortex solutions in an Abelian Chern-Simons theory with spontaneous breaking of gauge symmetry. Our equations above, without generalized function $H$ and with the parameter $l=1$, reproduces the BPS fundamental equations of the their article. 

\section{A particular case}

At this point, it is worth mentioning that the interest in the study of compacton topological vortices is due to the fact that applications of these configurations are found in many branches of physics, from condensed matter physics to nuclear physics \cite{AGW}. Thinking about it, when investigating these settings, we can remark that the $H(|\phi|)$ function is not specified explicitly. This way, let us treat two distinct possibilities for the generalization function $H(|\phi|)$ that lead us to vortices topological solutions. Initially, we consider that
\begin{equation}
    H=|\phi|^{2l}(|\phi|^{2l}-1)^2=F^{2l}g(r)^{2l}(1-F^{2l}g(r)^{2l})^2,
\end{equation}
where $l$ is a positive integer, and is responsible for making the scalar field more massive.

Thus the vortex equations are rewritten as
\begin{equation}
\label{b1}
    rg'(r)=\pm\frac{F^{l}}{r}a(r)g(r)^{l+1}(1-F^{2l}g(r)^{2l}),
\end{equation}
and
\begin{equation}
\label{b2}
    a'(r)=\mp\frac{m_{V}^{2}}{2}rg(r)^{2(l+1)}(g(r)^{2l}-1).
\end{equation}

In order to investigate the topological solutions of the model, we will assume the following boundary conditions:
\begin{align}
\label{boundary}
\nonumber
    g(0)=0, \hspace{2cm} g(\infty)=\nu \\
    a(0)=n, \hspace{2cm} a(\infty)=0,
\end{align}
where $\nu$ is the value of the vacuum state and $n$ is a defined positive integer parameter.

The behavior of $g(r)$ and $a(r)$ near vacuum is determined by solving the self-dual equations (\ref{b1}) and (\ref{b2}) around the boundary conditions (\ref{boundary}). Therefore, for $r\to 0$, the field profiles behave as
\begin{align}
    g(r)\simeq C_n r^n+...
\end{align}
meanwhile,
\begin{align}
    a(r)\simeq \frac{m_{V}^{2}}{2}r^2(C_n r^n)^{2(l+1)}\bigg\{\frac{1}{2[1+n(1+l)]}-\frac{(C_n r^n)^{(2l)}}{2[1+n(1+2l)]}\bigg\}+...
\end{align}
is numerically calculated. In this condition, when $0\leq g(r)<1$, both the generalization function $H(\vert\phi\vert)$ and the potential term of the model tend to have null contributions when we reach a compact-like field configuration, i. e., $l\to \infty$. On the other hand, let us analyze the asymptotic behavior of the model around $r\to \infty$. In this case, we have
\begin{align}
    g(r)\approx \nu+...
\end{align}
and
\begin{align}
    a(r)\simeq -\frac{m_{V}^{2}}{4}r^2\nu^2(\nu^{2l}-1)+...
\end{align}

It is observed that the coupling $H(|\phi|)$ and the source $B$ have minimal contributions (or null) when the parameter $l\to \infty$. However, it is important to mention that the source term $B$ has a significant contribution to the existence of the BPS properties of the model,  since the source is directly related to the spacetime curvature at the energy saturation limit. In this way, the field configurations are actually solitons with a profile compact-like and not true compactons (that have the boundary condition $g(r_c)=\nu$ with $\nu$ the vacuum of the theory). However, a slight modification of the theory's potential can be performed to obtain the  true compactons of the theory. For example, take the following potential 
\begin{align}\label{p}
    \mathcal{U}=\frac{e^2}{8}\beta(F^{2l}-\phi^{2l})^2.
\end{align}

Then, field settings described by true compactons appeared \cite{CLS}. In this case, near to the vacuum values, the potential will tend to a Heaviside function, generating field configurations that are described by compactons. This will be due to the fact that the boundary conditions will be rewritten as
\begin{align}\label{cc}
    g(r\to 0)=0, \, \, \, \, \, and \, \, \, \, \, g(r)=\nu
\end{align}
with $0\leq r<\infty$. However, we emphasize that the main difference between true compact and compact-like solutions will appear in the energy density and magnetic field behavior of the vortex. For example, for the case (\ref{p}), the compacton assumed an energy profile type a step function. On the other hand, compact-like solutions have a more localized profile as we will show next. 

It is important to make clear in our discussion that true compacton are configurations that reach the vacuum state in a finite region and that obey the condition (\ref{cc}). However, we will focus our attention on the compact-like solitons that obey the boundary conditions (\ref{boundary}).

\subsection{Numerical result}

\begin{figure}[h!]
\centering
\includegraphics[scale=0.7]{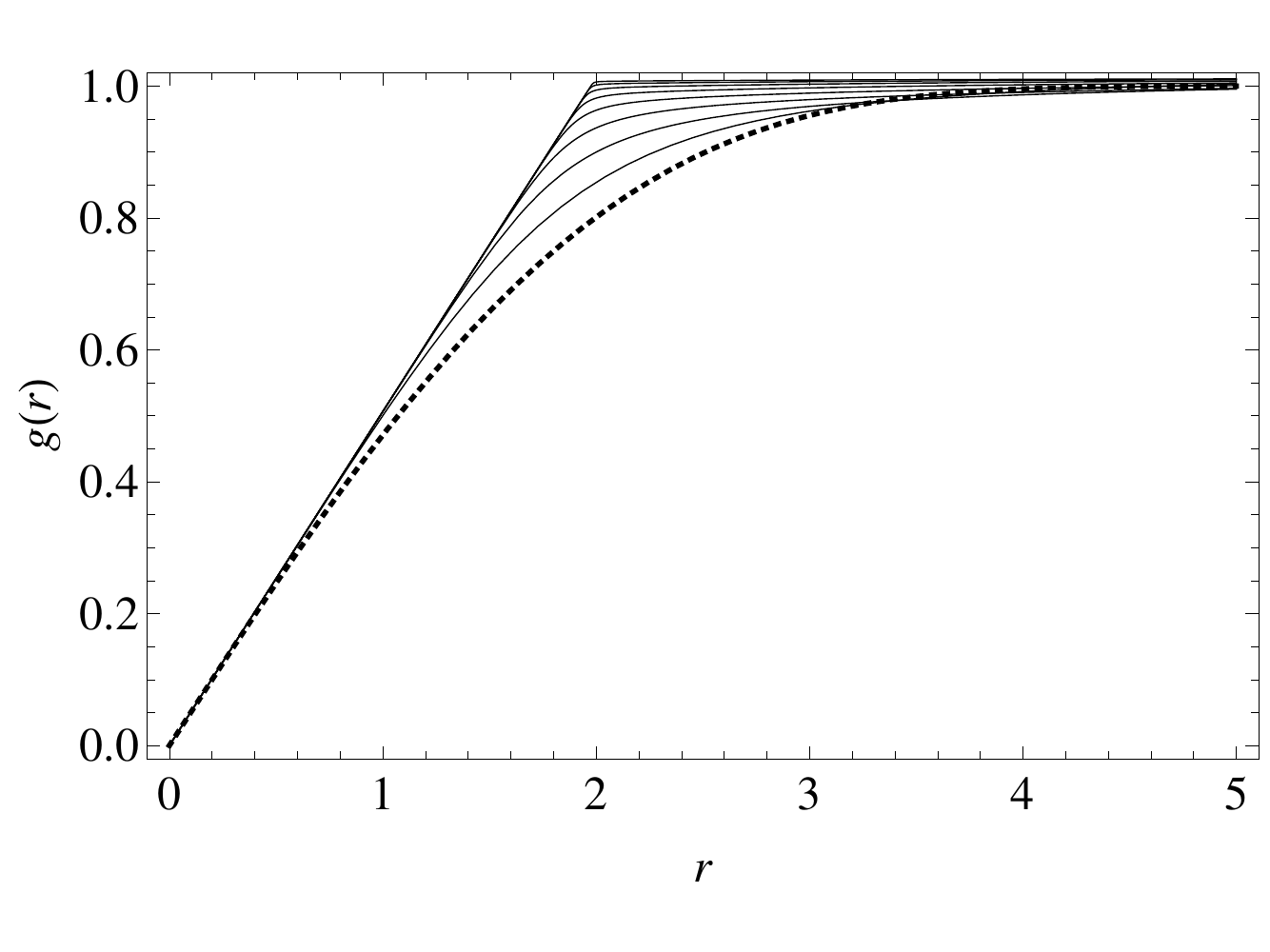}
\vspace{-20pt}
\caption{The numerical solutions of the scalar field for $l=1,2,4,8,16,32,64,128$ and $256$, with the case $l=1$ as the dotted line.} \label{fig1}
\end{figure}

\begin{figure}[h!]
\centering
\includegraphics[scale=0.7]{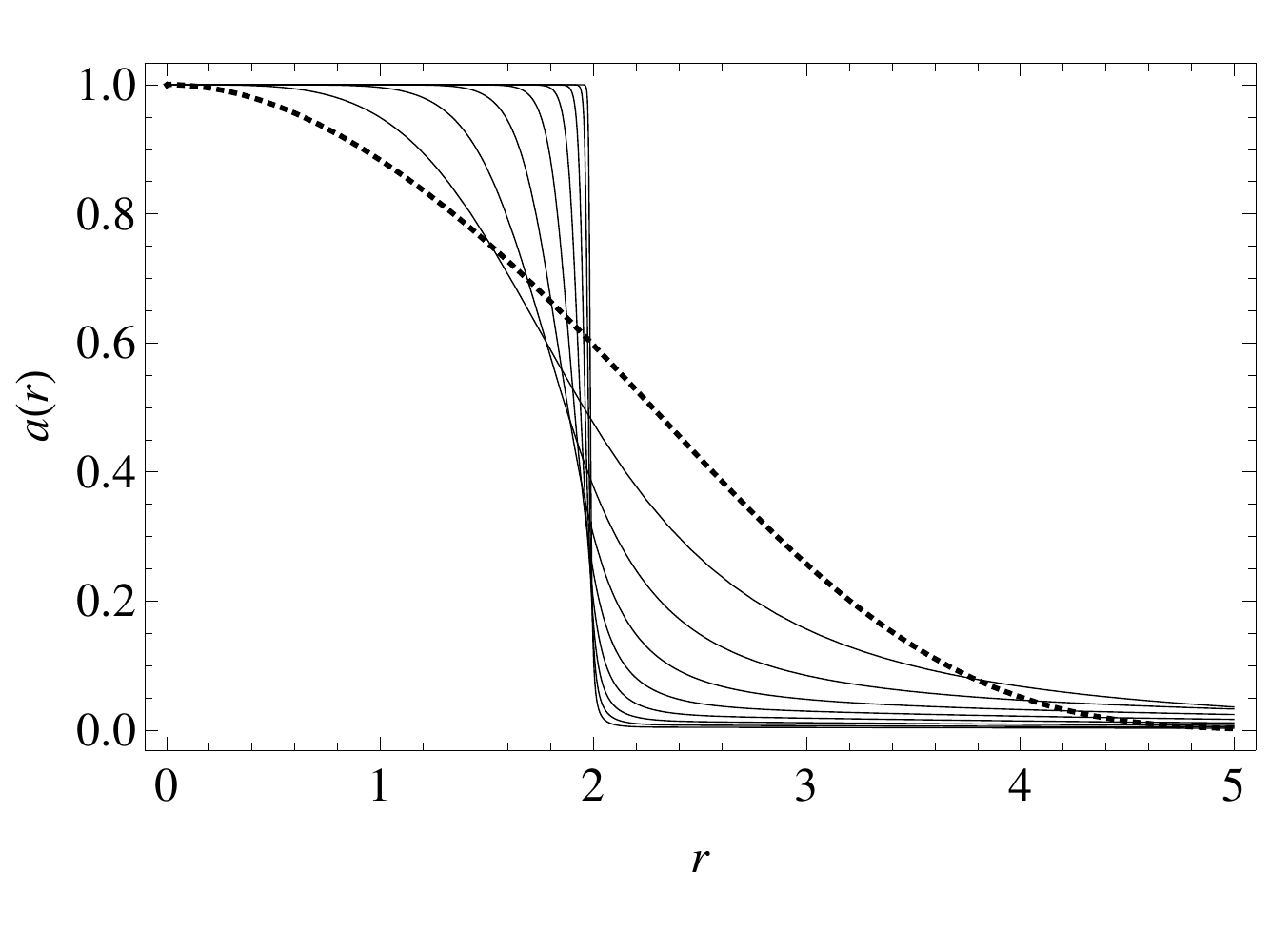}
\vspace{-20pt}
\caption{The numerical solutions of the gauge field for $l=1,2,4,8,16,32,64,128$ and $256$, with the case $l=1$ as the dotted line.} \label{fig2}
\end{figure}

At this point, we turn our attention to the numerical study of equations (\ref{b1}) and (\ref{b2}). For this, we use the boundary conditions expressed in eq. (\ref{boundary}) and through an interpolation we find the numerical solutions for the sector $0\leq |\phi|\leq 1$, shown in Fig. (\ref{fig1}). Then, from the expression (\ref{b1}) we obtain the numerical solution for the gauge field, shown in Fig. (\ref{fig2}).

\begin{figure}[h!]
    \centering
    \includegraphics[scale=0.7]{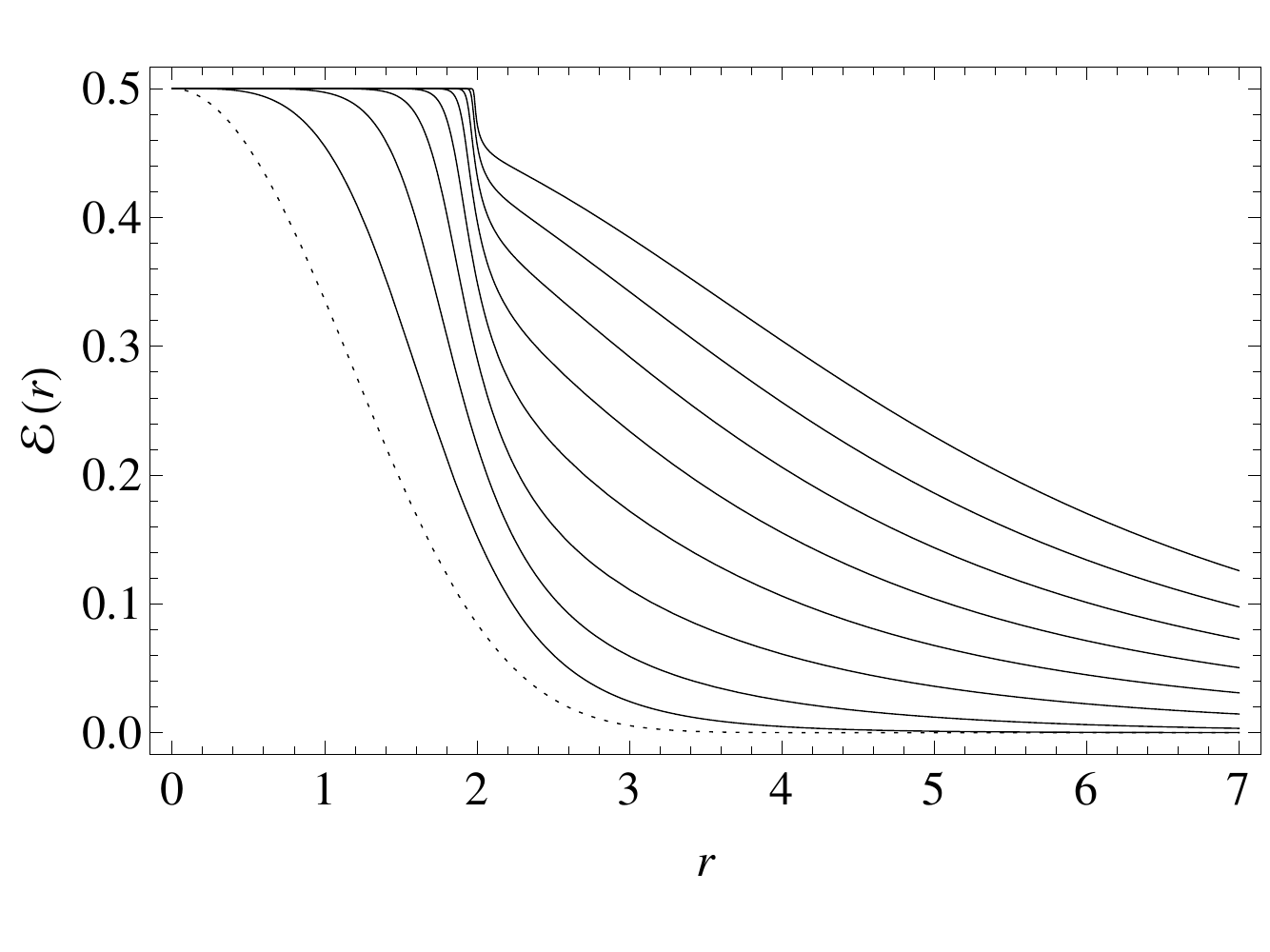}
    \vspace{-20pt}
    \caption{The energy of the model for $l=1,2,4,8,16,32,64,128, 256$ and $528$, where $l=1$ is the dotted line}.
    \label{fig3}
\end{figure}

Note that the solutions for the studied boundary give us a new class of topological solutions as was discussed in Ref. \cite{lima10}. As a consequence, the behavior of the energy density can be obtained from the compact-like solutions of the scalar field and the gauge field. Indeed, this can be accomplished through the expression (\ref{energy}). We present the result in Fig. (\ref{fig3}). We note that, when $g(r_c)=\nu$ the energy density presents a decay similar to an exponential curve.

\begin{figure}[h!]
    \centering
    \includegraphics[scale=0.7]{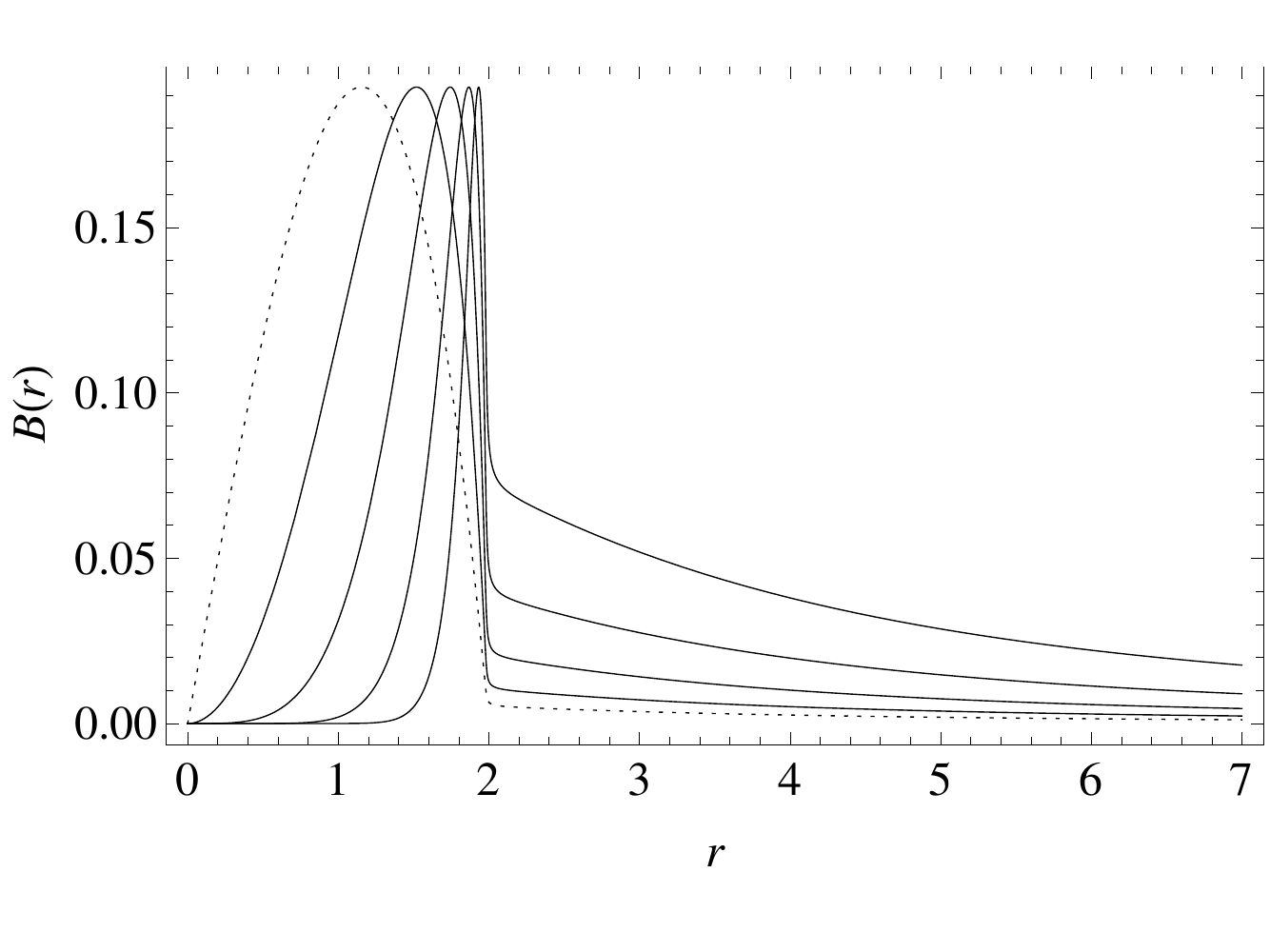}
    \vspace{-20pt}
    \caption{The magnetic field of the model for $l=1,2,4,8$ and $16$, where $l=1$ is the dotted line}.
    \label{fig4}
\end{figure}

Using the expression (\ref{mag}), we found the numerical solution for the magnetic field, which is sketched in Fig. (\ref{fig4}). We notice that it has a shape located around the value $r_c$, where $r_c$ is such that $ g (r_c) =\nu$.

This behavior of the emergence of compact-like structures in the model is a consequence of the fact that as the vacuum state becomes more localized, the Higgs field and the gauge field tend to stabilize quickly. In this way, the solutions reach the vacuum state in a short interval, resulting in structures known in the literature as compact-like solutions \cite{Bazeia}. It is interesting to mention that the emergence of such structures in the space-time $(3+1)D$ was never obtained. In this way, we realized that although we are working in a $(3+1)D$ scenario, our model behaves like the compact $(2+1)D$ vortex.


\section{Concluding remarks}

In this work we readdress the issue of vortex solutions of an Abelian Maxwell-Higgs theory. However, here the model is defined in a (3+1) dimensions curved space-time. Also, the scalar kinetic term is generalized by a function of the scalar field and is introduced a parameter that promote a new behavior for the vortices. Namely, for some values of the parameter we obtain vortices with compact features. Noteworthy, the curvature seems to work as an unexpected way. Our BPS equations reproduce, if desconsidering the generalization additions, the equations of an Abelian Chern-Simons vortices model (3D). Furthermore, for a particular function $H(\phi)$, we solve numerically the BPS equations and we observe that the gauge field assumes a characteristic similar to the step function. Also, it is important to note that for large $l$ values the energy presents a significant exponential decay from the value of $r_{c}$. As a consequence, we have a magnetic field more localized when $r=r_c$.
  

\section{Acknowledgments}
The authors thank the Conselho Nacional de Desenvolvimento Cient\'{\i}fico e Tecnol\'{o}gico (CNPq), grant n$\textsuperscript{\underline{\scriptsize o}}$ 308638/2015-8 (CASA), and Coordena\c{c}ao de Aperfei\c{c}oamento do Pessoal de N\'{\i}vel Superior (CAPES) for financial support. The authors thank the anonymous referee for his valuable criticisms and suggestions.



\begin{thebibliography}{150}

\bibitem{Nilsen}
H. B. Nilsen and P. Olesen, Nuc. Phys. B {\bf 61} (1973) 45.

\bibitem{Vega}
H. J. Vega and F. A. Schaposnik, Phys. Rev. D {\bf 14} (1976) 1100.

\bibitem{Jackiw}
R. Jackiw and E. Weinberg, Phys. Rev. Lett. {\bf 64} (1990) 2334. 

\bibitem{Singh}
V. Singh, D. A. Browne and R. W. Haymaker, Phys. Lett. B {\bf 306} (1993) 115.

\bibitem{Giacomo}
A. D. Giacomo, B. Lucini, M. Montesi and G. Paffuti, Nuc. Phys. B {\bf 73}, (1999) 524.

\bibitem{Creutz}
M. Creutz, Phys. Rev. D {\bf 21} (1980) 2308.

\bibitem{Haymaker}
R. W. Haymaker and J. Woziek, Phys. Rev. D {\bf 36} (1987) 3297.

\bibitem{BPS}
E. B. Bogomol'nyi, Sov. J. Nucl. Phys. {\bf 24} (1976) 449.

\bibitem{sch}
J. D. Edelstein, C. Nunez and F. A. Schaposnik, Phys. Lett. B {\bf 329} (1994) 39.

\bibitem{kimm1} K. Kimm, K. Lee, and T. Lee, Phys. Rev. D {\bf 53} (1996) 4436.

\bibitem{Ghosh}
P. K. Ghosh and S. K. Ghosh, Phys. Lett. B {\bf 366} (1996) 199.

\bibitem{Sales}
F. S. A. Cavalcante, M. S. Cunha and C. A. S. Almeida, Phys. Lett. B {\bf 475} (2000) 315.

\bibitem{dionisio} D. Bazeia, E. da Hora, C. dos Santos, and R. Menezes, Phys. Rev. D {\bf 81} (2010) 125014.

\bibitem{sour} L. Sourrouille, Phys. Rev. D {\bf 87} (2013) 067701.

\bibitem{LA}
F. C. E. Lima, and C. A. S. Almeida, Europhys. Lett. \textbf{131}, 31003 (2020).

\bibitem{LPA}
F. C. E. Lima, A. Yu. Petrov, and C. A. S. Almeida, Phys. Rev. D \textbf{103}, 096019 (2021).

\bibitem{bazeia} D. Bazeia, R. Casana, E. da Hora and R. Menezes, Phys. Rev. D {\bf 85} 125028 (2012).

\bibitem{casana2} R. Casana, M. L. Dias and E. da Hora, Phys. Lett. B {\bf 768} (2017) 254

\bibitem{lima10} F. C. E. Lima, D. M. Dantas, C. A. S. Almeida, Europhys. Lett. {\bf 130} (2020) 10005.

\bibitem{nam} J. Lee and S. Nam, Phys. Lett. B {\bf 261} (1991) 437.

\bibitem{bazeia2} D. Bazeia, Phys. Rev. D {\bf 46} (1992) 1879.

\bibitem{Picon} C. Armendariz-Picon, T. Damour and V. Mukhanov, Phys. Lett. B  {\bf 458}, (1999) 209.

\bibitem{Picon1} C. Armendariz-Picon, V. Mukhanov and P. J. Steinhard, Phys. Rev. Lett. {\bf 85}, (2000) 4438. 

\bibitem{rose} P. Rosenau and J. Hyman, Phys. Rev. Lett. {\bf 70} (1993) 565.

\bibitem{aro} H. Arodz, Acta Phys. Polon. B {\bf 33} (2002) 1241.

\bibitem{santos} C. dos Santos, Phys. Rev. D {\bf 82} (2010) 125009.

\bibitem{adam2} C. Adam, N. Grandi, P. Klimas, J. Sanchez-Guillen, and
A. Wereszczynski, J. Phys. A {\bf 41} (2008) 375401.

\bibitem{adam3} C. Adam, N. Grandi, P. Klimas, J. Sanchez-Guillen, and
A. Wereszczynski, Gen. Relativ. Gravit. {\bf 42} (2010) 2663.

\bibitem{bazeia3} D. Bazeia, L. Losano, M.A. Marques and R. Menezes, Phys. Lett. B {\bf 736}  (2014) 515.

\bibitem{diego1} D. F. S. Veras, W. T. Cruz, R. V. Maluf and C. A. S. Almeida, Phys. Lett. B {\bf 754} (2016) 201.

\bibitem{diego2} D. F. S. Veras and C. A. S. Almeida, Phys. Rev. D {\bf 95} (2017)104032.

\bibitem{lima1} F. C. E. Lima, D. A. Gomes and C. A. S. Almeida, Ann. Phys. {\bf 422} (2019) 168315.

\bibitem{T}
D. Tong and K. Wong, JHEP {\bf 01}, (2014) 090.

\bibitem{AQRW}
C. Adam, K. Oles, J. Queiruga, T. Romanczukiewicz, A. Wereszczynski, JHEP {\bf 1907}, (2019) 150.

\bibitem{S}
B. Schroers, SciPost Phys. {\bf 7}, (2019) 30.

\bibitem{NOB}
C. Naya, K. Oles, Phys. Rev. D {\bf 102}, (2020) 025007.

\bibitem{ASHCRO}
J. Ashcroft and S. Krusch, Phys. Rev. D {\bf 101}, (2020) 025004.

\bibitem{Edelstein}
J. D. Edelstein, G. Lozano and F. A. Schaposnik, Mod. Phys. Lett. A {\bf 8} (1993) 3665.

\bibitem{AGW}
C. Adam, J. Sanchez-Guillen, and A. Wereszczynski, Phys . Lett. B {\bf 691} (2010) 105.

\bibitem{gibbons1} A. Comtet and G. GIbbons, Nucl. Phys. B {\bf 299} (1988) 719.

\bibitem{gibbons2} G. Gibbons,  Lect. Notes Phys. 383 (1991) 110.

\bibitem{Coleman}
S. Coleman, S. Parke, A. Neveu and C. M. Sommerfield, Phys. Rev. D {\bf 15} (1977) 554.

\bibitem{Belavin}
A. A. Belavin, A. M. Polyakov, A. S. Schwartz and Yu S. Tyupkin, Phys. Lett. B {\bf 59} (1975) 85.

\bibitem{Bazeia}
D. Bazeia, L. Losano, M. A. Marques and R. Menezes, Phys. Lett. B, v. {\bf 772}, (2017) 253.

\bibitem{CLS}
R. Casana, G. Lazar, and L. Sourrouille, Adv. in High Energy Phys. \textbf{2018}, 1-20 (2018).


\end{thebibliography}
\end{document}